\documentclass[a4paper]{article}

\usepackage{amsfonts}
\usepackage{graphicx}

\begin{document}

\title{$1/f$ Noise in Fractal Quaternionic Structures}

\author{\textbf{T.\,Me\v{s}kauskas}$^{1,2}$ and \textbf{B.\,Kaulakys}$^{2}$}

\date{
\textit{
$^1$Vilnius University, Naugarduko 24, LT-03225 Vilnius, Lithuania\\[0.5\baselineskip]
$^2$Institute of Theoretical Physics and Astronomy of Vilnius University,
Go\v stauto 12, LT-01108 Vilnius, Lithuania\\[0.5\baselineskip]
}
Email:
\texttt{tadas$\!$.$\!$meskauskas@maf$\!$.$\!$vu$\!$.$\!$lt}
}

\maketitle

\begin{abstract}
We consider the \textit{logistic map} over \textit{quaternions}
$\mathbb{H}\sim\mathbb{R}^4$ and different 2D projections of
Mandelbrot set in 4D quaternionic space. The approximations (for
finite number of iterations) of these 2D projections are fractal
circles. We show that a \textit{point process} defined by radiuses
$R_j$ of those fractal circles exhibits pure $1/f$ noise.


\end{abstract}

\section{Introduction}

$1/f$ noise is observed in large diversity of real life and artificial systems, which behavior is usually defined by a complex interaction
of many components. Complexity of the system usually assumes  that long-term correlations are observed.
Examples are processes and experimental data in condensed matter, traffic flow, quasar emissions, music, biological and medical systems,
economic and financial data, human cognition and even distribution of prime numbers (see \cite{Pilgram} and references herein).

Fluctuations of signals defined by time series obtained from such
systems are found to be cha\-rac\-te\-rized by a \textit{power
spectral density} $S(f)$ diverging at low frequencies $f$ like
$1/f^\alpha$, here $\alpha$ is some real parameter. $1/f$
($\alpha\approx 1$) noise is an intermediate between the white
noise ($\alpha=0$) with no correlation in time and the random walk
(Brownian motion) noise ($\alpha=2$) with no correlation between
increments. Note that Brownian motion can be obtained integrating
white noise and that taking the integral of the signal increases
the exponent $\alpha$ by 2 while the inverse operation of
differentiation decreases it by 2.

Parameter $\alpha$ is closely related to the Hurst exponent $H$.
It is known that fluctuations which are fractionally homogeneous,
i.e. unifractal or uniscaling, can be quantified by a single coefficient $H$
and a single exponent $\alpha$ \cite{Peng}.

Possible generalization leads to multiscaling or multifractals,
with the exponent H dependant on time. Therefore multifractal processes are characterized
by a set of scaling relations or power laws with correspondingly many exponents $\alpha$ \cite{Mandelbrot}.

\section{Point processes and $1/\lowercase{f}$ noise}

In many cases, the intensity of some current can be represented by
a sequence of random (however, as a rule, mutually correlated)
or pseudo-periodic pulses $A_k(t-t_k)$. Here the function
$A_k(\varphi)$ represents the shape of the $k$-th pulse having an influence
to the current $I(t)$ in the region of transit time $t_k$.
The intensity of the current in some space cross-section may,
therefore, be expressed as
$$
   I(t) = \sum_k A_k (t-t_k).
$$

It is easy to show that the shapes of the pulses mainly influence the high
frequency, $f\ge \Delta t_p$ with $\Delta t_p$ being the characteristic
pulse length, power spectral density of $I(t)$ while fluctuations
of the pulse amplitudes result, as a rule, in the white or Lorentzian
but not $1/f$ noise.

\begin{figure}[h]
  \begin{center}
  \includegraphics[clip=true, width=125mm]{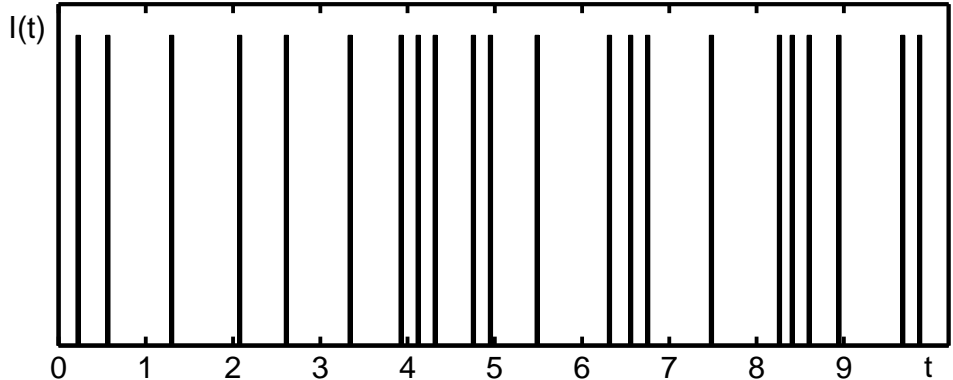}
  \end{center}
  \caption{Current $I(t)$ vs time $t$ defined by (\ref{I_apibrezimas}) formula. Such dependencies appear when registering the consecutives heart beats,
  cars on a highway passing through the reference point, transactions in financial markets etc.}
  \label{fig_I}
\end{figure}

Therefore, we restrict our analysis to the fluctuations due to the
correlations between the transit times $t_k$ and hence we can
replace the function $A_k(t-t_k)$ by the Dirac delta function
$\delta (t-t_k)$. The current (see Fig.\,\ref{fig_I}) is then
expressed as
\begin{equation}
   I(t) = \sum_k \delta (t-t_k).
   \label{I_apibrezimas}
\end{equation}

Following this approach, instead of current $I(t)$, we further deal with \textit{point process}, defined
by the sequence $t_1, t_2, \ldots, t_N, \ldots $.

The \textit{power spectral density} of the current
(\ref{I_apibrezimas}) is defined as
\begin{equation}
   S(f) = \lim_{N\to\infty}
   \frac{2}{t_N} \left|
   \sum_{k=1}^N e^{\displaystyle -i2\pi ft_k} \right|^2
   \label{Sf}
\end{equation}
where $[t_1,t_N]$ is assumed to be the interval of observation.

In this approach the power spectral density of the signal depends on the
statistics and correlations of \textit{point process} (the transit times $t_k$) only.
It is well known that sequence of random, Poisson, transit times
generates white (shot) noise, for example.

In \cite{PRE} we proposed simple analytically solvable model for producing \textit{point process}
resulting in $S(f)\sim 1/f$ ($\alpha=1$) noise. Discussion on the origin and universality
of $1/f$ noise was continued in \cite{MR,ICNF99}. Some further work, related to the
applications of the theory of \textit{point processes} and $1/f$ noise to econophysics, was done in \cite{GK1,GK2}.

\section{Quaternions and other hypercomplex numbers}

\textit{Complex numbers} $\mathbb{C}\sim\mathbb{R}^2$, along to their \textit{real} predecessors $\mathbb{R}$, are widely
used in nowadays mathematical modeling and scientific computing. Beside others, they have important applications in theories of complex systems,
fractals and signal processing: famous Mandelbrot and Julia fractal sets are defined in $\mathbb{C}$, spectrum (Fourier transform) is
defined as integral of complex function etc.

There are some clues that we should not stop with the computations in $\mathbb{R}$ and $\mathbb{C}$, and that further
generalization to \textit{quaternions} $\mathbb{H}\sim\mathbb{R}^4$ (introduced by Hamilton) or even \textit{octonions} $\mathbb{G}\sim\mathbb{R}^8$
(introduced by Graves) are particulary interesting and valuable, even though the role of these \textit{hypercomplex} numbers
is not widely understood yet.

In order to define \textit{hypercomplex} algebras, one has to consider not only
two algebraic operations $+$ and $\times$, but also one geometric map:
$x \mapsto \bar x$, where $\bar x$ denotes the conjugate vector of $x$.

The three operations are defined recursively as we define the algebras,
in the following manner.
Let $A_k$ be the real \textit{hypercomplex} algebra of dimension $2^k$, $k\ge 1$.
It is constructed recursively as $A_k = A_{k-1}\times A_{k-1}$ by means
of the three following operations:
\begin{eqnarray*}
   \lefteqn{\mbox{addition:}}\hphantom{\mbox{multiplication:}}\quad
   (a,b)+(c,d) & = & (a+c,b+d),\\
   \lefteqn{\mbox{conjugacy:}}\hphantom{\mbox{multiplication:}}\quad
   \lefteqn{\overline{(a,b)}}\hphantom{(a,b)+(c,d)} & = & (\bar{a},-b),\\
   \mbox{multiplication:}\quad
   (a,b) \times (c,d) & = & (ac - \bar{d} b, da + b \bar{c}),
\end{eqnarray*}
where $ac$ denotes $a\times c$ in $A_{k-1}$.
For $k=0$, $A_0$ is taken to be the field $\mathbb{R}$ with the arithmetic
operations $+$ and $\times$, the conjugacy map being the identity on $\mathbb{R}$:
$a\mapsto \bar a=a\in\mathbb{R}$. This construction is known to algebraists
as the Cayley-Dickson doubling process.

About computations with \textit{hypercomplex} numbers, and why only \textit{real numbers}, \textit{complex numbers},
\textit{quaternions} and \textit{octions} are suitable for computations see \cite{CERFACS,NA} and references herein.

Explicitly multiplication in $\mathbb{H}$ can be expressed as
$(a,b,c,d)\times (a',b',c',d') = (a'',b'',c'',d'')$, with
\begin{eqnarray*}
      a'' = aa' - bb' - cc' - dd'\\
      b'' = ab' + ba' + cd' - dc'\\
      c'' = ac' + ca' + db' - bd'\\
      d'' = ad' + da' + bc' - cb'
\end{eqnarray*}

\section{$1/\lowercase{f}$ noise in quaternionic Mandelbrot set}

We consider the {\textit{logistic map} over \textit{quaternions} $\mathbb{H}\sim\mathbb{R}^4$
\begin{equation}
    z_{k+1} = r z_k (1-z_k),  \quad r, z_k \in \mathbb{H}, \quad  k = 0, 1, \ldots .
    \label{Logistines}
\end{equation}
with given initial value $z_0$, for example $z_0=(0.5,0,0,0)$. The
logistic map (\ref{Logistines}) has been extensively studied over
$\mathbb{R}$ (real numbers) and $\mathbb{C}$ (complex numbers).
Despite its great simplicity this map exhibits an extremely
complex behaviour. The study of (\ref{Logistines}) on $\mathbb{R}$
gives birth to the Feigenbaum tree while the analysis of
(\ref{Logistines}) on $\mathbb{C}$ leads to the famous Mandelbrot
and Julia fractal sets.

Further we deal with 2D projections of Mandelbrot set in 4D quaternionic
space. Any two components of $r$ are set to zero, while the
remaining two vary. For example,
$$
    {\cal{M}}_{12} = \left\{ (r_1,r_2): r = (r_1,r_2,0,0), \lim_{k\to\infty} |z_k| < \infty
    \right\},
$$
$$
    {\cal{M}}_{24} = \left\{ (r_2,r_4): r = (0,r_2,0,r_4), \lim_{k\to\infty} |z_k| < \infty
    \right\}.
$$
Note that ${\cal{M}}_{12}$ is just the famous Mandelbrot set in
$\mathbb{C}$. We also show that
${\cal{M}}_{12}={\cal{M}}_{13}={\cal{M}}_{14}$ and
${\cal{M}}_{23}={\cal{M}}_{24}={\cal{M}}_{34}$.

\begin{figure}[h]
  \begin{center}
  \includegraphics[clip=true, height=50mm]{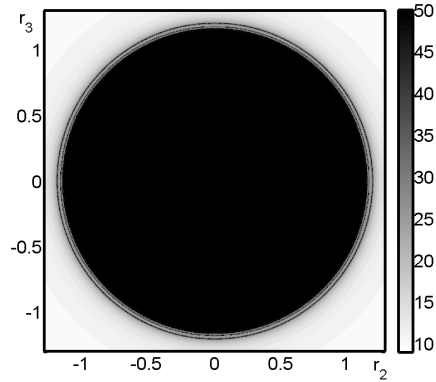}
  \end{center}
  \caption{Approximation (after $50$ ite\-ra\-tions) of Mandelbrot set ${\cal{M}}_{23}$ (one gets
  exactly the same for ${\cal{M}}_{24}$ or ${\cal{M}}_{34}$).}
  \label{fig_M23}
\end{figure}

The approximations (for finite number of iterations) of Mandelbrot
set ${\cal{M}}_{23}={\cal{M}}_{24}={\cal{M}}_{34}$ (near its
boundary) are fractal circles (see Fig.\,\ref{fig_M23}), dependant
only on radius $R=\sqrt{r_2^2+r_3^2}$.

\begin{figure}[h]
  \begin{center}
  \includegraphics[clip=true, width=120mm]{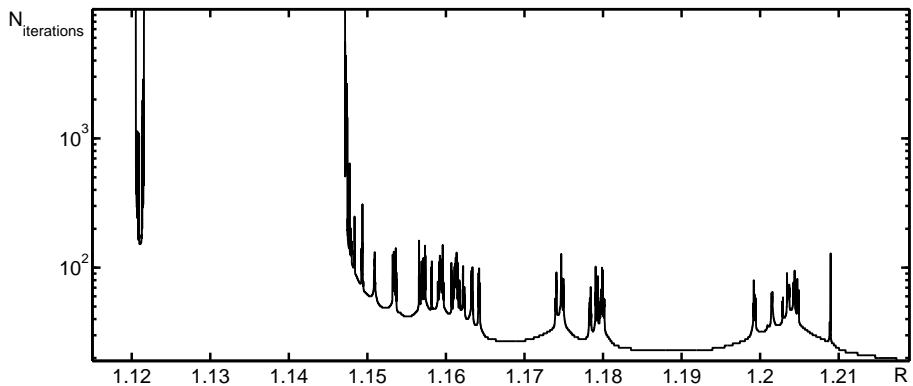}
  \end{center}
  \caption{The number of iterations needed to reach $|z_k|>10^{10}$ plotted vs radius $R$ when computing ${\cal{M}}_{23}$.}
  \label{fig_NumbIt}
\end{figure}

\begin{figure}[h]
  \begin{center}
  \includegraphics[clip=true, width=120mm]{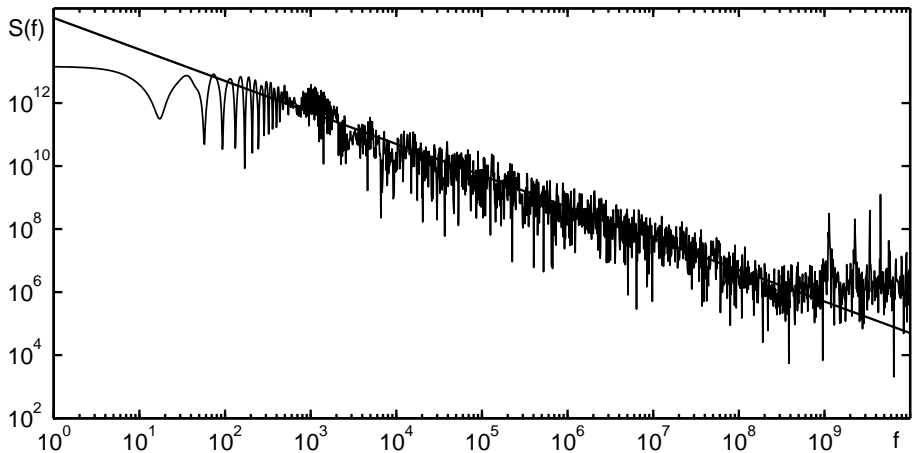}
  \end{center}
  \caption{Power spectral density $S(f)$, defined by (\ref{SpektrasR}), vs frequency $f$ with $N=796474$. The plot is compared to the function $1/f$.}
  \label{fig_Sf}
\end{figure}

Define the \textit{point process} $R_j$ as the values of radius of
each circle -- mathematically they are the values of $R$, small
change of which result in significant change of number of
iterations needed for $|z_k|$ to reach ``infinity'' ($10^{10}$ for
example). The values $R_j$ correspond to peaks in
Fig.\,\ref{fig_NumbIt}}.

According to (\ref{Sf}), the \textit{power spectral density} of
such \textit{point process} is defined as
\begin{equation}
  S(f) \approx \frac{2}{R_N-R_1} \left|\, \sum_{j=1}^N e^{\displaystyle -i 2\pi f R_j}
  \,\right|^2,
  \label{SpektrasR}
\end{equation}
here $N$ is the volume of point process data ($N\to\infty$ as $R_j$ recording resolution increases).

We obtain (see Fig.\,\ref{fig_Sf}) that $S(f)\sim 1/f$,
\textit{i.\,e.} radiuses $R_j$ of fractal circles in Mandelbrot
set ${\cal{M}}_{23}$ exhibit pure $1/f$ noise ($\alpha=0$) or
unifractal noise.

\section*{Acknowledgments}

We acknowledge support by the Lithuanian State Science and Studies Foundation.

\end{document}